\documentclass[aps,prl,amsmath,amssymb,floatfix,twocolumn, showpacs, superscriptaddress]{revtex4}
\usepackage{multirow}
\usepackage{bbold}
\usepackage{graphicx}
\usepackage{subfigure}
\usepackage{color}
\usepackage{hyperref}

\usepackage{amsfonts, relsize, color}
\usepackage{graphics}
\usepackage{graphicx}
\usepackage{subfigure}
\usepackage{hyperref}
\usepackage{color}

\begin{document}
\title{Note on ``Diffusive quantum criticality in three-dimensional disordered Dirac semimetals"}

\author{Bitan Roy}
\affiliation{Condensed Matter Theory Center and Joint Quantum Institute, University of Maryland, College Park, Maryland 20742-4111, USA}

\author{S. Das Sarma}
\affiliation{Condensed Matter Theory Center and Joint Quantum Institute, University of Maryland, College Park, Maryland 20742-4111, USA}

\date{\today}

\begin{abstract}
We correct erroneous conclusions from Ref.~\cite{roy-dassarma} regarding the values of various critical exponents, calculated to the two-loop order, and argue that $\epsilon$-expansion near two spatial dimension, with $\epsilon=d-2$ may not be reliable to address the critical properties of disorder driven Dirac semimetal-metal quantum phase transition in $d=3$.  
\end{abstract}


\maketitle

We here clarify some errors in our previous two-loop calculation for potential disorder driven Dirac semimetal (DSM)-compressible diffusive metal (CDM) quantum phase transition (QPT) in three spatial dimensions ($d=3$), presented in Ref.~\cite{roy-dassarma}. In Ref.~\cite{roy-dassarma}, we performed a two-loop analysis within the framework of an $\epsilon$-expansion near $d=2$ (the lower critical dimension for the problem), and reported that the correlation length exponent (CLE) ($\nu$) is given by $\nu^{-1}=\epsilon-\frac{\epsilon^2}{8}+{\mathcal O}(\epsilon^3)$ and dynamic scaling exponent (DSE) $z=1+\frac{\epsilon}{2}-\frac{3}{16}\epsilon^2+{\mathcal O}(\epsilon^3)$, where $\epsilon=d-2$, near the DSM-CDM quantum critical point (QCP). Hence, for $d=3$, we found $\nu \approx 1.14$ and $z \approx 1.31$~\cite{roy-dassarma}. We now realized that the quoted values of these two exponents are incorrect, because we missed some relevant two-loop diagrams, which we correct here. But, more importantly we here question the validity of such an $\epsilon$-expansion near $d=2$ and argue that the $\epsilon$-expansion may not be reliable for this problem.

We realized that instead of reevaluating the diagrams shown in Ref.~\cite{roy-dassarma}, one can derive the flow equation for disorder coupling ($\Delta$) by analyzing the Gross-Neveu (GN) model for a \emph{discrete} chiral symmetry breaking mass condensation for which the flow equation of GN coupling ($g$) is known to three-loop order~\cite{gracey, luperini-rossi, kivel-stepanenko, moshe-moshe}. The Euclidean actions for these two theories are respectively
\begin{eqnarray}
\bar{S}_D&=&\int d^d \vec{x} d \tau \; \left(\bar{\Psi}_{a} \left[-i \partial_\tau \gamma_0-i v \gamma_j \partial_j \right] \Psi_{a}\right)_{\tau} \nonumber \\
&-&\frac{\Delta}{2} \int d^d \vec{x} d \tau d \tau^\prime \left(\bar{\Psi}_{a} \gamma_0 \Psi_{a} \right)_{\tau} \left(\bar{\Psi}_{b} \gamma_0 \Psi_{b} \right)_{\tau^\prime},\label{action-disorder} \\
S_{GN}&=& \int d^D \vec{x} \; \left[ \bar{\Psi} \left(-i \gamma_\mu \partial_\mu \right)\Psi-g \left( \bar{\Psi} \Psi\right)^2 \right], \label{action-GN} 
\end{eqnarray}   
where $\bar{\Psi}=\Psi^\dagger \gamma_0$, $a,b$ are replica indices. Due to fermion doubling in a lattice, we work with four dimensional $\gamma$ matrices, satisfying the anticommutaiton relation $\left\{ \gamma_\mu, \gamma_\nu\right\}=2 \delta_{\mu \nu}$. Critical behavior of $S_{GN}$ and $\bar{S}_{D}$ can be studied performing $\epsilon$-expansions respectively near two space-time and spatial dimensions. The infra-red flow equation for $g$ to three-loop order reads as~\cite{gracey, luperini-rossi, kivel-stepanenko, moshe-moshe} 
\begin{equation}\label{GN-threeloop}
\beta_g=-\epsilon g + (n-2) g^2 -(n-2)g^3-(n-2)(n-7)\frac{g^4}{4},
\end{equation}
where $\epsilon=D-2$ with $D$ as \emph{space-time} dimensions, and $n$ is the number of spinor components.

\begin{figure}[htb]
\centering
\includegraphics[width=8.0cm, height=4.cm]{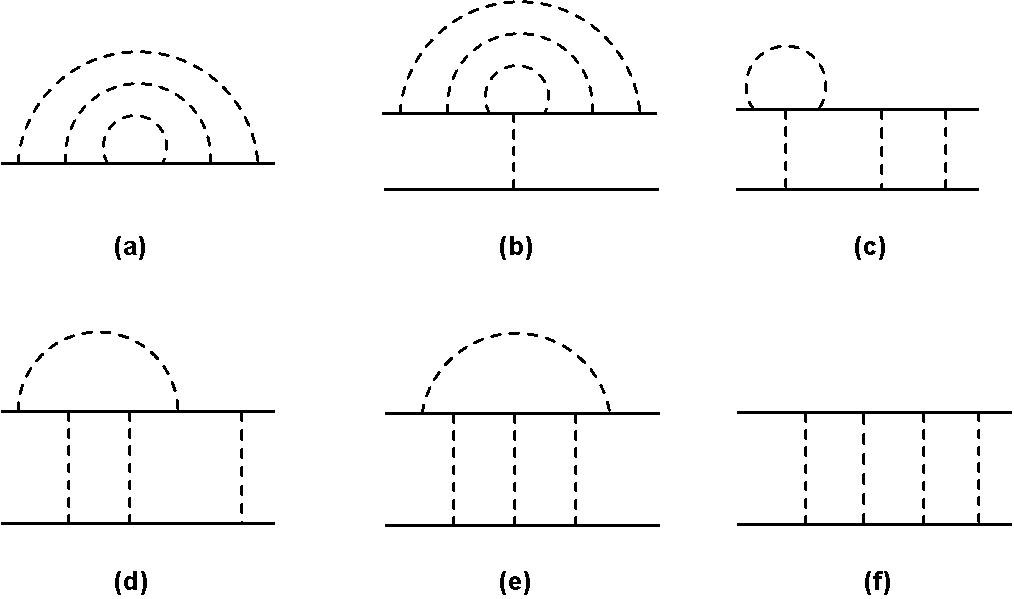}
\caption{Three loop diagrams contributing to the flow of $\Delta$. We here do not show various combination of each class of diagrams, obtained by exchanging the disorder (dashed) lines. Notice three-loop diagram (f) and its various combinations, and two-loop diagrams $(viii)$, $(ix)$, $(xi)$, $(xii)$ from Ref.~\cite{roy-dassarma} do not renormalize $\Delta$ or $g$, since $\gamma_j \gamma_k \gamma_l \sim \epsilon_{jklm}\gamma_5 \gamma_m$.}
\label{three-loop}
\end{figure}

To derive the flow equation of $(\Delta)$ one can set the external frequency to zero, for which we establish a correspondence between diagramatic contributions for $S_{GN}$ and $\bar{S}_D$. Contributions from  relevant (i) one-loop (Fig.~$(i)$-$(iv)$ of Ref.~\cite{roy-dassarma}) and three-loop [see Fig.~\ref{three-loop}] diagrams are of opposite sign when the fermion vertex is accompanied by the matrix $\mathbb{1}_{4 \times 4}$ (GN theory) and $\gamma_0$ (potential disorder), (ii) two-loop diagrams that renormalize either GN ($g$) or disorder ($\Delta$) coupling are of same sign for these two theories. Thus we can arrive at the flow equation for $(\Delta)$ to three-loop order from Eq.~(\ref{GN-threeloop}) by (i) first setting $n=0$, since any contribution proportional to $n$ involves fermion bubble that vanishes in the vanishing replica limit, and (ii) taking $-g=\Delta$, leading to  
\begin{equation}\label{flow-3loop}
\beta_{\Delta}=-\epsilon \Delta+2 \Delta^2+2\Delta^3 +\frac{7}{2} \Delta^4,
\end{equation} 
where $\epsilon=d-2$ and $d$ represents the \emph{spatial} dimensions. The exact location of the DSM-CDM QCP to from one-, two and three-loop order are respectively $\Delta^{\ast}_{(1)} = \epsilon/2$, $\Delta^{\ast}_{(2)}=\left(-1+\sqrt{1+2 \epsilon} \right)/2$ and 
\begin{equation}
\Delta^{\ast}_{(3)}=-\frac{4}{21}+ \sqrt{\frac{272}{441}} \sinh \left[ \frac{1}{3} \sinh^{-1} \left[ \frac{ 1323 \epsilon + 440}{136\sqrt{7}}\right] \right]. \nonumber
\end{equation}
When we expand the last two quantities in powers of $\epsilon$ the radius of convergence is $1/2$ and $\leq 0.35$ (extracted numerically). Thus, setting $\epsilon=1$ to extract the CLE in $d=3$ is not a well defined procedure, since this expansion is not asymptotically convergent. Therefore, higher order terms from the $\epsilon$-expansion about two spatial dimensions only produces unreliable results for $\epsilon=1$ and this methodology of addressing the critical properties of DSM-CDM QPT in $d=3$ needs to be abandoned.     

Even if we choose to extract CLE from the powers series in terms of $\epsilon$, to one-, two- and three-loop orders we respectively find $\nu^{-1}_{(1)}=\epsilon+{\mathcal O}(\epsilon^2)$, 
\begin{eqnarray}\label{CLE}
\nu^{-1}_{(2)}=\epsilon+\frac{\epsilon^2}{2}+{\mathcal O}(\epsilon^3), \:
\nu^{-1}_{(3)}=\epsilon+\frac{\epsilon^2}{2}+ \frac{3}{8}\epsilon^3+{\mathcal O}(\epsilon^4). \nonumber
\end{eqnarray}
Therefore, in three spatial dimensions ($\epsilon=1$), we obtain $\nu_{(1)}= 1$, $\nu_{(2)}\approx 0.67$ and $\nu_{(3)} \approx 0.53$. The fact that the inclusion of higher order corrections pushes the CLE close to the mean-field value ($\nu=1/2$) is alarming, indicating failure of $\epsilon$-expansion to capture the critical properties. It may be worthwhile to mention that the higher loop results for the CLE either do not ($\nu_{(3)}$) or barely ($\nu_{(2)}$) satisfy the Chayes-Chayes-Fisher-Spencer inequality $(\nu>2/d)$~\cite{CCFS}, while the one-loop result $(\nu=1)$ satisfies the inequality, indicating that going beyond the one-loop order may not be meaningful in this problem.

Furthermore, one may wish to capture the asymptotic behavior of CLE using the \emph{Pad\'{e} approximation}~\cite{pade, comment-borel}. For $\nu^{-1}_{(2)}$ a Pad\'{e} $[1|1]$ approximation yields
\begin{equation}\label{pade11}
\left[\nu^{-1}_{(2)}\right]^{[1|1]}=\frac{\epsilon}{1-\frac{\epsilon}{2}}.
\end{equation}  
Consequently, we obtain a mean-field value of CLE $\nu_{(2)}=1/2$ for $\epsilon=1$. On the other hand, to estimate the asymptotic behavior of $\nu^{-1}_{(3)}$ we can employ either Pad\'{e} $[2|1]$ or Pad\'{e} $[1|2]$ approximations, respectively yielding 
\begin{eqnarray}
\left[\nu^{-1}_{(3)}\right]^{[2|1]}=\frac{\epsilon-\frac{\epsilon^2}{4}}{1-\frac{3}{4}\epsilon}, \quad
\left[\nu^{-1}_{(3)}\right]^{[1|2]}=\frac{\epsilon}{1-\frac{\epsilon}{2}-\frac{\epsilon^2}{8}}, 
\end{eqnarray}  
in turn giving $\nu_{(3)} \approx 0.33$ and $0.375$ for $\epsilon=1$, both being smaller than the mean-field value of CLE, strongly suggesting the inapplicability of $\epsilon$-expansion for $d=3$. 

We do not discuss the corrections to the DSE ($z$) beyond the leading order in $\epsilon$. Given that the radius of convergence of $\Delta^{\ast}_{(2,3)} \ll 1$, we conclude that higher order corrections to $z$ are not meaningful when $\epsilon=1$.

Therefore, within the $\epsilon$-expansion scheme, where $\epsilon=d-2$, only reliable values of the critical exponents near DSM-CDM QCP are $\nu=\epsilon^{-1}$ and $z=1+\epsilon/2$, respectively giving $\nu=1$ and $z=3/2$ in three spatial dimensions ($\epsilon=1$), which one can also obtain from one-loop Wilsonian renormalization group analysis~\cite{goswami}. Going to higher loop calculations in $\epsilon$ for $d=3$ can only produce progressively worse results for the critical exponents since the corresponding series is manifestly not convergent. 

We point out that recent numerical calculations for the DSE ($z$) agree with the result from one-loop $\epsilon$-expansion within numerical error bars~\cite{herbut, brouwer, pixley, ohtsuki, roy-bera}, whereas the one-loop result for CLE ($\nu$) reasonably agrees (within error bar) with the numerically calculated exponents in Ref.~\cite{herbut, pixley, ohtsuki, roy-bera}, but disagrees with Ref.~\cite{brouwer}. Whether the agreement between the numerical results and the one-loop calculation is merely a coincidence or implies something deep about the problem is unknown at this stage.

B. R. is thankful to John Gracey for valuable communication related to perturbative analysis of Gross-Neveu model. B. R. also thanks Igor Herbut and Pallab Goswami for useful discussion on this problem. During the preparation of the erratum, we became aware of a preprint~\cite{syzranov} where a two-loop calculation has been carried out explicitly. Our flow equation in Eq.~(\ref{flow-3loop}) to two-loop order agrees with the one reported in Ref.~\cite{syzranov}.


\begin{thebibliography}{10}
\bibitem{comment-borel} Since the power series of CLE does not display oscillatory behavior we cannot perform a Borel summation.
\end{thebibliography}
\end{document}